\documentclass[prl,twocolumn,showpacs]{revtex4}
\usepackage{array,amsmath,amssymb,graphicx}

\begin{document}
\title{Bose-Einstein Condensates in Strongly Disordered Traps}

\author{T. Nattermann}
\affiliation{Institut f\"ur Theoretische
Physik, Universit\"at zu K\"oln,
Z\"ulpicher Str. 77, D-50937 K\"oln,
Germany}

\author{V.L. Pokrovsky }
\affiliation{Department of Physics, Texas
A\&M University, College Station, Texas
77843-4242}  \affiliation{Landau Institute
for Theoretical Physics, Chernogolovka,
Moscow District, 142432, Russia}

\date{\today}

\begin{abstract}
A Bose-Einstein condensate in an external
potential consisting of a superposition of
a harmonic and a random potential is
considered theoretically. From a
semi-quantitative analysis we find the
size, shape and excitation energy as a
function of the disorder strength. For
positive scattering length and
sufficiently strong disorder the
condensate decays into  fragments each of
the size of the Larkin length ${\cal L}$.
This state is stable over a large range of
particle numbers. The frequency of the
breathing mode scales as $1/{\cal L}^2$.
For negative scattering length a
condensate of size ${\cal L}$ may exist as
a metastable state. These findings are
generalized to anisotropic traps.
\end{abstract}

\maketitle

   Bose-Einstein condensation (BEC) is
of great interest in a wide variety of
systems including superfluidity and
superconductivity. Its perhaps cleanest
realization is found in
 dilute ultracold gases of alkali atoms
confined in magnetic or optical traps
which have been studied in great detail
both experimentally \cite{Ketterle,Wieman}
and theoretically \cite{Dalvovo99,Leggett}
(and references therein).

More recently these investigations have
been extended to traps formed by  a
superposition of a harmonic and a random
potential \cite{Clement05,Lye05,Fort05}.
Whereas for harmonic traps the behavior of
the condensate, e.g. its size, shape,
elementary excitations etc. is quite well
understood, this is not the case for traps
including a finite amount of disorder.
Most of the theoretical investigations of
BEC in random potentials use as a starting
point a description in momentum space
(e.g. Bogoliubov transformation or
Beliaev-Popov perturbation theory) where
the study of the typical disorder effects
is notoriously difficult
\cite{Huang,Pitaevskii,Vinokur,Pelster}.
Disorder is there treated  perturbatively
with $\xi/{\cal L}$ as a small parameter,
where $\xi$ and ${\cal L}$ denote the
healing and the Larkin length,
respectively. As we will show below in the
case of sufficiently strong disorder this
parameter is of order one and perturbation
theory breaks down (but see
\cite{Graham}).

In the present article we present a
semi-quantitative analysis of BEC on the
level of Larkin-Imry-Ma arguments
\cite{Larkin_70,Imry} to determine the
size, shape and the  frequencies of the
breathing mode in weakly interacting
Bose-Einstein condensates in disordered
traps. In addition to the oscillator  and
the scattering length the Larkin length
appears as a new relevant length scale
which dominates  the properties of the
condensate for weak interaction and strong
disorder. Strong disorder reduces the
anisotropy of the condensate. For
attractive interaction the condensate may
exist in a metastable state.


\emph{The model.}
Starting point is the Hamiltonian of
interacting bosons
\begin{equation}
{\cal H} = \int d^3x\Psi^{\dag}\Big
(-\frac{\hbar^2}{2m}{\nabla}^2 +
U(\textbf{x}) +
\frac{2\pi\hbar^2a}{m}\Psi^{\dag}\Psi\,\Big)\Psi.
\label{eq:hamiltonian}
\end{equation}
Here $\Psi^{\dag}(\textbf{x})$ and
$\Psi(\textbf{x})$ are the creation and
annihilation operators of the Bose field
with $\int d^3x|\Psi|^2=N$ for the
particle number. $m$ denotes the mass of
the bosons and $a$ the scattering length.
The external potential represents a
superposition of the harmonic trap and a
Gaussian random potential
\begin{equation}\label{}
U(\textbf{x})=\frac{m}{2}
\sum_{i=1}^3\omega_i^2x_i^2+U_{\textrm{dis}}(\textbf{x}).
\end{equation}
The random  potential is  characterized by
\begin{equation}\label{} \langle
U_{\textrm{dis}}\rangle=0 ,\,\,\,\,\,
\langle
    U_{\textrm{dis}}(\textbf{x})U_{\textrm{dis}}(\textbf{x}')\rangle=
    {\kappa^2}\delta(\textbf{x}-\textbf{x}').
\end{equation}
With this ansatz we assume that the
correlation length of the disorder is
smaller than all other length scales. We
will briefly discuss other cases at the
end of this article.

In what follows we consider the state of
the system at zero temperature. We will
first examine the case of an isotropic
$3$-dimensional trap with
$\omega_1\approx\omega_2\approx\omega_3=\omega_{\textrm{o}}$.
Assuming a compact condensate cloud of the
radius $R$ the condensate energy per
particle $\varepsilon$  can be written as
\begin{equation}\label{eq:energy}
    \varepsilon(R)\approx
    \frac{\hbar^2}{2m}
    \left(\frac{1}{R^2}+3\frac{aN}{R^3}+
    \frac{R^2}{\ell^4}
    -\frac{4}{3(R^{3}{\cal L})^{1/2}}\right).
\end{equation}
Here we introduced the oscillator length
(the size of the harmonic oscillator
ground state)
$\ell=(\hbar/(m\omega_{\textrm{o}}))^{1/2}$.
${\cal L}$ denotes the Larkin length
${\cal
L}=16\pi\hbar^4/(27m^2\kappa^2)\quad$
\cite{Larkin_70,Imry} in $d=3$ dimensions.
$N$ is the total particle number. In the
following we will set $\hbar=m=1$. In this
notation the healing length in the absence
of disorder reads $\xi=\sqrt{R^3/(6Na)}$.

The first and the second term in
(\ref{eq:energy}) describe the kinetic
energy and the interaction of the
particles, respectively. Both terms favor
the spreading of the condensate provided
the scattering length is positive. If the
radius of the cloud is sufficiently large,
namely
\begin{equation}\label{eq:interaction}
    R \gg R_a=3aN,
\end{equation}
the interaction is negligible in
comparison with the kinetic energy.
Neglecting the inhomogeneity of the
condensate, condition
(\ref{eq:interaction}) can be rewritten as
$\xi\gg R$.  The third and the fourth term
describe the oscillator potential and the
typical potential well from the disorder,
respectively, both cause its confinement.
In the last term we have taken into
account that the typical value of the
fluctuation of the potential in a region
of linear size $R$ scales as $\kappa
R^{-3/2}$. This term is larger than the
kinetic energy only on scales
$R\gtrsim{\cal L}$. Similarly to the
interaction, at sufficiently large radius
of the cloud,
\begin{equation}\label{eq:disorder}
     R \gg R_{\cal L}=\ell (\ell/{\cal
     L})^{1/7},
\end{equation}
the disorder can be neglected in
comparison  to the harmonic potential.

In general the center of an attractive
domain formed by static fluctuations of
the random potential  may not coincide
with the center of the harmonic trap.
Nevertheless, our estimates are correct
since it is either the harmonic trap or
the disorder which leads to the
localization of the condensate.
Experimentally localization by disorder is
often observed by a sudden decrease of the
oscillator frequency. In this case the
condensate will be localized in the coarse
grained potential well closest to the
origin.


\emph{{Condensate size}.}
Below we  determine the equilibrium size
$R_0$ of the condensate
 as a function of
the particle number $N$ and the disorder
strength  from the equilibrium condition
$\partial \varepsilon/\partial R=0$. In
the case of weak disorder and relatively
small particle number, such that
 $ \ell\ll {\cal L}$ and $N\ll\ell/(3a)$, we can ignore both the
 disorder and the interaction and hence,
 from (\ref{eq:energy}) $R_0\approx \ell$.
 At large particle number $N>\ell/(3a)$, the
 effect of the interaction sets in and
 $R_0$ approaches the Thomas-Fermi radius
 \cite{Timmermanns1997}
\begin{equation}\label{eq:ThomasFermiradius}
R_{\textrm{TF}}(N)
    \sim \big(aN\big)^{1/5}
    \ell^{4/5}\,
\quad    .
\end{equation}
 More interesting
is the case of strong disorder, such that
the oscillator length is much larger than
the Larkin length, $\ell \gg {\cal L}$.
 Then, for small particle number, the competition
 between the kinetic energy
and the effective potential well resulting
from the random potential on scale $R$
gives
\begin{equation}\label{eq:newradius1}
    R_0\approx {\cal L}\,,  \quad Na\ll {\cal
    L}.
\end{equation}
Thus the size of the condensate is equal
to the Larkin length.  We estimate the
healing length as $\xi\sim \sqrt {{\cal
L}^3/Na}\gg {\cal L}$ from which it is
plausible to conclude that the condensate
is strongly depleted and the superfluid
stiffness vanishes \cite{Huang}.

For larger particle numbers the
interaction becomes important and the
minimum of (\ref{eq:energy}) is given by
\begin{equation}\label{eq:newradius2}
    R_0(N)\approx \Big(\frac{9}{2}Na\Big)^{2/3}{\cal
    L}^{1/3}\,,\quad  {\cal
    L}\ll aN\ll \ell\Big(\frac{\ell}{\cal L}\Big)^{5/7}.
\end{equation}
The (average) healing length is estimated
from $\xi\sim \sqrt{R_0^3/Na}\sim({\cal
L}/Na)^{1/6}R_0\ll R_0$  from which we
concludes that the condensate is weakly
depleted only and the superfluid stiffness
remains finite \cite{Huang}.
 At the upper
boundary for $N$ the radius reaches
$R_{\cal L}$. Finally, for even larger
particle number the radius crosses over to
the Thomas-Fermi radius. The crudeness of
our energy expression (\ref{eq:energy})
does not
 allow  accurate determination of the numerical pre-factors
for  all quantities derived from
(\ref{eq:energy}). If the disorder
decreases, the Larkin-length ${\cal L}$
increases and the interval where
(\ref{eq:newradius2}) applies diminishes
and eventually vanishes when ${\cal L}$
reaches the oscillator length $\ell$. In
this case only the cross-over at
$Na\approx \ell$ discussed previously
survives.


\emph{Fragmentation.}
Next we consider the instability of a
compact condensate against fragmentation.
For $N\ll {\cal L}/a$ the ground state
energy for increasing $N$ decreases  as
$E\approx -N/(6{\cal L}^2)$ until it
reaches a plateau $E\approx E_0$ at
$N\gtrsim N_0\approx {\cal L}/a$ where
$E_0\approx -{2}/({27}{\cal L}a)$ (compare
(\ref{eq:newradius2})). Thus, at $N>N_0$,
the division of the condensate into
$N/N_0$ fragments is energetically
favorable. The energy of the fragmented
condensate is
$E\approx(N/N_0)E_0(N_0)\approx-N/{\cal
L}^2<E_0(N)$.
 Thus the fragmented condensate
corresponds to the true ground state. The
total volume of all fragments scales like
$\sim Na{\cal L}^2$ and is hence by a
factor $({\cal L}/(Na))$ smaller than the
volume of the compact state. The healing
length $\xi$ in the fragments is
$\xi=({\cal L}^3/(6Na))^{1/2}$ where
$N\lesssim N_0$, and hence $\xi\gtrsim
{\cal L}$. According to \cite{Huang}, in
the region where $\xi\gtrsim {\cal L}$
 a finite fraction of the bosons are not
 in the condensate. The superfluid stiffness of
 the fragmented state  most likely vanishes.
\begin{figure}[h,t,b]
\includegraphics[width=.40\linewidth]{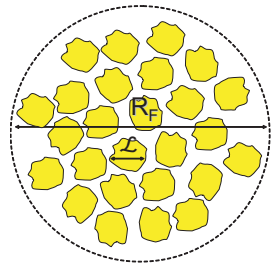}
\caption{State of dense fragments. Each
fragment includes ${\cal L}/a$ particles.
The minimal condensate radius is $R_F\sim
(Na)^{1/3}{\cal L}^{2/3}$. } \label{QD}
\end{figure}
\begin{figure}[h,t,b]
\includegraphics[width=0.70\linewidth]{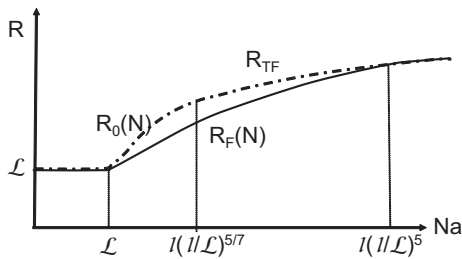}
\caption{Condensate radius as a function
of the particle number. The bold line
corresponds to the equilibrium fragmented
state, the dashed line corresponds to the
metastable compact state.} \label{QD}
\end{figure}
To describe the cross-over to the
Thomas-Fermi behavior we consider a
 state of dense fragments with the total
 radius $R_F\sim (Na)^{1/3}{\cal L}^{2/3}$. $R_F$ denotes the minimal size of
 of the fragmented state. If there is no confining parabolic potential
the fragments may be distributed over a
much larger area to take advantage of the
tails of the Gaussian distribution of
disorder. In a fragmented state with
radius $R_F$ the
 oscillator energy per particle is  of the order
 $R_F^2\ell^{-4}$ which has to be
 compared with the energy per particle $-{\cal L}^{-2}$ from the
 disorder. This gives for the
 cross-over particle number
 \begin{equation}\label{eq:cross-over}
    N_{co}a\sim {\ell}{} \left({\ell}/{\cal
    L}\right)^5.
 \end{equation}
This number is much larger then the
particle  cross-over number between
(\ref{eq:newradius2}) and
(\ref{eq:ThomasFermiradius}) provided
$\ell>{\cal L}$.

The compact state described by
(\ref{eq:newradius2})  may occur as a
metastable state which lives  for a short
period of time after a confining harmonic
trap is switched off (or the oscillator
frequency $\omega_o$ is strongly reduced).
On larger time scales the condensate will
lower its energy  by decaying into $\sim
Na/{\cal L}$ fragments each of the size
${\cal L}$.  This decay into fragments may
explain the striped phases found in
cigar-like traps \cite{Lye05}.


\emph{Breathing mode.}
Next we consider
the breathing mode frequency $ \omega_B$
of the radial oscillations of the
condensate around the equilibrium
configuration from
\begin{equation}\label{eq:oscillations}
m\omega_B^2=\varepsilon^{\prime\prime}(R_0).
\end{equation}
Without disorder we find from
(\ref{eq:energy}) and
(\ref{eq:oscillations})
$\omega_B=2\omega_{\textrm{o}}$ in the
non-interacting and $\omega={\sqrt
5}\omega_{\textrm{o}}$ in the interacting
case, respectively, in agreement with
previous results \cite{Stringari}.

 In the disordered case we obtain
for $Na\ll{\cal L}\ll\ell$
\begin{equation}\label{}
    \omega_B \approx
    \frac{1}{\sqrt 2}
    \left(\frac{\ell}{{\cal L}}\right)^2\omega_{\textrm{o}}
    \gg \omega_{\textrm{o}}
\end{equation}
and for ${\cal L}\ll
aN\ll\ell({\ell}/{\cal
    L})^{5/7}$ under the assumption of a
    compact non-fragmented condensate
\begin{equation}\label{}
\omega_B\sim \frac
{\ell^{2}}{(Na)^{7/6}{\cal
L}^{5/6}}\,\omega_{\textrm{o}}\gg
\omega_{\textrm{o}}.
\end{equation}
Contrary to the pure case the interaction
changes the radial oscillations of the
condensate strongly.

In the true equilibrium fragmented state
the breathing mode will be controlled by
the weak interaction between the
fragments. In this case the frequency gap
will probably decrease as $\sim R_F^{-2}$.


 \emph{Negative scattering length.}
So far we assumed that the scattering
length $a$ is positive. In the case of
negative scattering length as e.g. in the
case of $\,^7$Li atoms \cite{Bradley},
there is only a metastable state of finite
radius $R_c$ following from $\partial
\varepsilon/\partial R|_{R=R_c}=0$. This
state becomes unstable at a critical
particle number $N_c$. The condition for
the disappearance of the metastable
minimum follows from
\begin{equation}\label{eq:criticalradius}
    \frac{\partial
\varepsilon(R)}{\partial
R}=\frac{\partial^2
\varepsilon(R)}{\partial
R^2}\Big|_{R=R_c}=0
\end{equation}
In the pure case ${\cal L}\gg \ell$,
condition (\ref{eq:criticalradius}) gives
for the critical particle number
    $N_c
    \approx
    0.12{\ell}/{|a|}$ in agreement with \cite{Bradley,Shuryak96,Kagan96,Kagan_98}.

The strong disorder decreases the radius
of the metastable droplet as it  follows
from the equilibrium condition
\begin{equation}\label{}
    \left(\frac{\cal
    L}{R}\right)^{1/2}-\frac{4}{27}\frac
    {N}{N_c}\left(\frac{\cal
    L}{R}\right)^{3/2}=1.
\end{equation}
Here $N_c$ denotes the critical particle
number beyond which the metastable state
disappears
\begin{equation}\label{eq:Ncriticaldisorder}
    N_c=\frac{8}{243}\frac{{\cal L}}{|a|}\approx
    0.033\frac{{\cal L}}{|a|}.
\end{equation}
The radius of the metastable droplet in
the region $N<N_c$ decreases monotonically
from $R_c\approx {\cal L}$ for $N\ll N_c$
to $R_c\approx \frac {2}{9}{\cal L}$ for
$N\approx N_c$.


\emph{Anisotropic traps}. So far we
assumed that the trap is isotropic whereas
experimentally often anisotropic  traps
are considered. Our results can be easily
extended to these cases. For brevity  we
will consider here only the strongly
anisotropic situations (i)
$\omega_1=\omega_2=\omega_{\textrm{o}}\ll
\omega_3=\omega_{{\perp}}$ ($d=2)$ and
(ii) $\omega_1=\omega_{\textrm{o}}\ll
\omega_2=\omega_3=\omega_{{\perp}}$
($d=1$)
 corresponding to a pancake and
a cigar-shape like trap, respectively. We
 assume that the the oscillator length
obeys
$\ell_{{{\perp}}}=(\hbar/m\omega_{{\perp}})^{1/2}\ll
\ell, {\cal L}$ so that the extension of
the condensate in the transverse direction
is given by $\ell_{{{\perp}}}$. The
extension $R_0$ in the longitudinal
direction follows then from the minimum of
the energy per particle
\begin{equation}\label{eq:energygeneral}
    \varepsilon\approx
\frac{1}{2R^2}+\frac{3aN}{2R^d\ell_{{{\perp}}}^{3-d}}+
    \frac{R^2}{2\ell^4}-
    \frac{2}{d(R^{d}
    {\cal L}_d^{4-d})^{1/2}},\quad d=1,2.
\end{equation}
Here we have introduced the $d-$dimensonal
Larkin length ${\cal L}_d=({\cal
L}\ell_{\perp}^{3-d})^{1/(4-d)}$ and
assumed that the correlation length $b$ of
the random potential is small compared
with the transverse width of the trap,
$b\ll \ell_{\perp}$. In the opposite case
$b\gg \ell_{\perp}$, $\ell_{\perp}$ has to
be replaced by $\ell_{\perp}^2/b$ in the
definition of ${\cal L}_d$. From
(\ref{eq:energygeneral}) we can read off
the generalized cross-over radii
\begin{equation}\label{}
    R_a\sim \ell_{{{\perp}}}
    \left(\frac{\ell_{\perp}}{aN}\right)^{1/(2-d)},\quad
    R_{\cal L}=\ell
    \left(\frac{\ell}{{\cal
    L}_d}\right)^{\frac{4-d}{4+d}}.
\end{equation}
For $R\gg R_a$ the interaction  and for
$R\gg R_{\cal L}$ the disorder can be
neglected, respectively. For a
\emph{d-dimensional} trap we find in
analogy to (\ref{eq:newradius1}) for small
particle numbers
\begin{equation}\label{newradius1,d}
{R_0}\approx {\cal L}_d\,, \quad Na
<\ell_{\perp}\left({\ell_{\perp}}/{{\cal
L}_d}\right)^{{2-d}}.
\end{equation}
{In this region the healing length
$\xi\gtrsim {\cal L}_d$ such that a finite
fraction of particles is not in the
condensate and superfluidity is
suppressed.} For larger $N$
(\ref{eq:newradius2}) has to be replaced
by
\begin{equation}\label{eq:newradius3}
{R_0}\approx
{\ell_{\perp}}\left({aN}/{\ell_{\perp}}
   \right)^{{2}/{d}}
   \left({{\cal L}_d}/{\ell_{\perp}}\right)^{(4-d)/d}.
\end{equation}
This formula is valid as long as
\begin{equation}\label{eq:newradius4}
    aN<\ell_{\perp}\left({\ell_{\perp}}/{\ell}\right)^{2-d}
    \left({\ell}/{{\cal
L}_d}\right)^{\frac{(4-d)(d+2)}{4+d}}
\end{equation}
For even larger $N$ one find the
generalized Thomas-Fermi behavior
\begin{equation}\label{}
{R_{TF}}(N)\approx{\ell_{\perp}}
\left({aN\ell^{4}}/{\ell_{{{\perp}}}^{5}}\right)^{\frac{1}{2+d}}
\end{equation}
However, we have to consider here again
the instability of the compact
 domain state described by (\ref{eq:newradius3}) against fragmentation.
The energy of the compact state scales as
$-(1/a\ell_{\perp})(\ell_{\perp}/{\cal
L}_d)^{4-d}$ whereas that of the
fragmented state scales as $-N/{\cal
L}_d^2$. Thus the fragmented state has the
lower energy until the radius of the set
of dense fragments
\begin{equation}\label{}
    R_F(N)\sim \ell_{\perp}(Na{\cal
L}_d^2/\ell_{\perp}^3)^{1/d}
\end{equation}
reaches a value such that ${\cal
L}_d^{-2}\sim R_F^2/\ell^4$ which happens
at
\begin{equation}\label{eq:cross-over-generalized}
    N_{co}\sim \left({\ell_{\perp}^{3-d}}/{\ell^{2-d}a}\right)
    \left({\ell}/{{\cal
    L}}\right)^{2+d}
\end{equation}
which generalizes (\ref{eq:cross-over}),
but is more realistic, especially at
$d=1$.

In conclusion, we demonstrated that the
disorder strongly influences the size,
shape and structure of the condensate
cloud. Sufficiently strong disorder leads
to the localization of the condensate
fragments of linear size of the Larkin
length. This result agrees with the
conclusions about the suppression of the
transport in elongated Bose-condensates
derived in \cite{Clement05}. The
increasing number of particles enhances
interaction and may delocalize the
condensate. Such a delocalization
transition is especially realistic in a
strongly prolonged cigar-shape condensate.
If the atoms in condensate attract each
other, the disorder decreases the critical
number of particles at which collapse is
developed.

This work has been supported by
DFG-project NA222/5-2 (T.N. and V.P.) and
by the DOE under the grant DE-FG02-06ER
46278 (V.P.).

Electronic mail:
\emph{natter@thp.uni-koeln.de} and
\emph{valery@physics.tamu.edu}.


\end{document}